\documentclass[amsmath,amssymb]{revtex4}
\usepackage{graphicx} 
\usepackage{color}
\definecolor{Red}{rgb}{1.0,0.0,0.0}

\begin{document}

\title{Indenting fractal-edged elastic materials}

\author{D. F. S. Costa}
\affiliation{Departamento de F\'{i}sica, Universidade Federal do 
Cear\'{a}, 60451-970 Fortaleza, Cear\'{a}, Brazil}

\author{J. H. M Pontes}
\affiliation{Departamento de F\'{i}sica, Universidade Federal do 
Cear\'{a}, 60451-970 Fortaleza, Cear\'{a}, Brazil}

\author{W. P. Ferreira}
\affiliation{Departamento de F\'{i}sica, Universidade Federal do 
Cear\'{a}, 60451-970 Fortaleza, Cear\'{a}, Brazil}

\author{J. S. de Sousa}
\affiliation{Departamento de F\'{i}sica, Universidade Federal do 
Cear\'{a}, 60451-970 Fortaleza, Cear\'{a}, Brazil}
  
\author{C. L. N. Oliveira}
\email{lucas@fisica.ufc.br}
\affiliation{Departamento de F\'{i}sica, Universidade Federal do 
Cear\'{a}, 60451-970 Fortaleza, Cear\'{a}, Brazil}

\begin{abstract}
Surface roughness plays a crucial role in the accuracy of indentation
experiments used to measure the elastic properties of materials. In
this study, we present a computational analysis of how surface
roughness, represented explicitly by fractal geometry, influences the
mechanical properties of soft materials. We model two-dimensional
elastic samples with a Koch fractal bottom surface, grown upward or
downward to the fourth generation, referred to as fractal
\textit{down} and fractal \textit{up}, respectively. The
elastodynamics equations are solved numerically while a rigid punch
indents the elastic sample from the top surface. By applying the Hertz
model for mechanical contact, we determine the Young's modulus of the
materials. Our findings reveal that fractal surfaces, especially those
with dimensions comparable to the sample size, can significantly alter
experimental measurement outcomes. In particular, the roughness of the
substrate profoundly affects the measured elastic properties, as seen
in scenarios involving cell elasticity. For instance, in the
\textit{down} fractal scenario, reductions in the measured elastic
modulus range from 2\% to 4\%, while increases reach up to 40\% in the
\textit{up} fractal scenario. These results underscore the importance
of incorporating fractal geometry into the design and analysis of
indentation experiments. This approach could significantly enhance our
understanding and application of material characterization and
mechanical testing, leading to more accurate and reliable results.
\end{abstract}

\maketitle

\section{Introduction}

Interfacial effects profoundly influence the behavior of virtually
every physical system~\cite{Allara2005}, ranging from quantum
devices~\cite{Oliveira2004a, Oliveira2004b} and chemical
compositions~\cite{Prutton1994} to transport
phenomena~\cite{Andrade2004, Oliveira2010, Lenzi2014}. These effects
can mislead estimations if not carefully considered, which is
especially critical in experiments involving surface interactions. One
common type of such experiment is indentation testing, where a hard
tip is pressed into a material's surface to measure penetration depth
and applied force, thereby evaluating mechanical
properties. Traditional models like the Hertz mechanical contact model
assume perfectly smooth surfaces and uniform stress distribution
throughout the material. However, real-world materials often deviate
from this ideal due to surface irregularities, such as microscale and
nanoscale roughness in metals and polymers,
respectively~\cite{Chuah2019}. These discrepancies highlight the need
for a nuanced approach to interpreting experimental data, considering
the complex realities of material surfaces.

Surface roughness disrupts the assumption of uniform stress
distribution, creating complex stress patterns that vary with the
material's topographical features at different scales~\cite{Yang2018,
  Mohammadi2012, McMillan2018}. These irregularities can significantly
alter experimental outcomes by inducing stress concentrations not
fully accounted for by standard models~\cite{Xiao2020,
  Tiwari2021}. Addressing these complexities requires advanced
modeling techniques and multi-scale mechanical characterization to
assess a material's properties~\cite{Peng2023} accurately. Moreover,
surface irregularities can compromise a material's yield strength,
lead to premature plastic deformation, and interfere with measurements
of elastic recovery, potentially resulting in incorrect assessments of
a material's ability to revert to its original shape
post-indentation~\cite{Tiwari2020}.

Furthermore, nanoindentation assays on living cells provide insights
into rheologic behaviors sensitive to their microenvironment and
internal physical properties~\cite{Sousa2017, Sousa2020, Lima2024,
  Moura2023, Discher2005}. The interaction of proteins like integrins,
which form focal adhesions connecting the cell to the substrate and
activate intracellular signaling pathways, illustrates this
sensitivity~\cite{Parameswaran2014}. Cells can deform the contact surface on soft substrates
such as the extracellular matrix, creating mechanical feedback that
influences cell adhesion further. Such irregular adhesion topographies
pose challenges and offer opportunities to explore cellular adaptation
and mechanobiology~\cite{Rianna2017}. Additionally, physiological
processes or external stimuli driving changes in the cytoskeleton can
significantly alter cell's mechanical properties~\cite{Wen2011,
  Trepat2008}, evident in diseases like cancer where such properties
serve as diagnostic markers~\cite{Rebelo2013, Lekka2016, Fisseha2012,
  Abidine2021}. Thus, understanding the influence of substrate
roughness on these measurements is crucial for developing more
effective diagnostic tools and enhancing our understanding of cellular
mechanics in health and disease.

\begin{figure}[t]
  \centering
  \includegraphics[width=0.9\columnwidth]{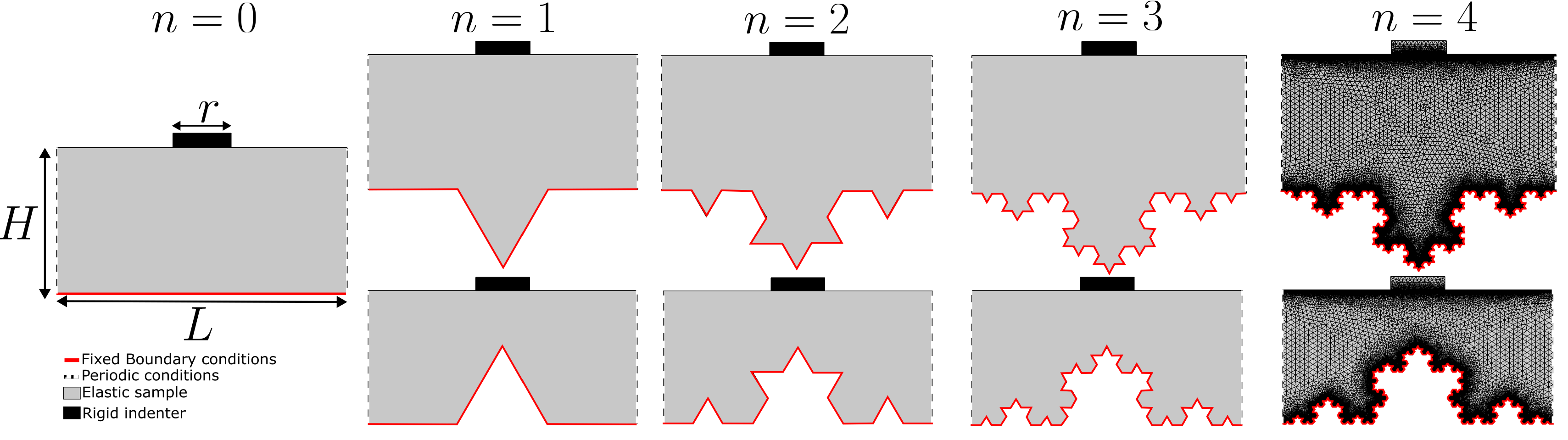}
  \caption{Illustration of the indentation physical model, featuring
    an elastic sample colored in gray, with height $H$ and length $L$,
    compressed from above by a rigid indenter of length $r=L/5$.  The
    top surface, shown in a black solid line, is free to displace,
    while the bottom surface, represented by a red solid line, remains
    fixed. Periodic boundary conditions are applied on the sides to
    simulate an infinite medium. The bottom surface transforms
    according to the rules of a Koch fractal. At the first generation
    ($n=0$), the sample maintains a rectangular shape without fractal
    growth. Subsequent frames visually demonstrate the progressive
    growth of the fractal pattern from $n=1$ to $n=4$, expanding
    either upwards or downwards. The numerical mesh for solving the
    elastodynamics equations is depicted for the ($n=4$) cases in up
    and down configurations.} \label{fig1}
\end{figure}

This study focuses on the impact of surface roughness, particularly
fractal-like roughness characterized by self-similar patterns across
scales, on the measurement of elastic properties. The presence of
fractal geometry in surface roughness introduces substantial
challenges, necessitating sophisticated models and analytical
techniques for accurate property evaluation. We utilize numerical
simulations to explore stress and strain distributions during
indentation on surfaces modeled as Koch fractals, helping us
understand how these patterns affect material behavior under stress
and enabling us to design experiments that more accurately reflect the
mechanical properties of materials with complex
surfaces~\cite{Bogahawaththa2024, Pavon2021, Wang2018}. Such insights
are critical in practical applications, especially in nanoindentation
tests on soft materials with rough substrates, where surface roughness
can directly affect the measurement of crucial parameters like
elasticity~\cite{Roy2024, Afferrante2023}. Understanding the
fundamental mechanics governing complex material behavior under
indentation is essential for advancing material design and selection
in engineering applications and developing more effective diagnostic
approaches in medical research~\cite{Sokolov2015, Baish2000,
  Sedivy1997, Dokukin2011}.

\section{Computational model}

Our model comprises a two-dimensional elastic medium characterized by
Young's modulus, $E_s$, a mass density, $\rho$, and a Poisson ratio,
$\nu$. The top surface of this medium is free to move, while the
lateral edges adhere to periodic boundary conditions to mitigate
finite-size effects. Conversely, the bottom surface is fixed, evolving
according to the $n^{th}$ generation of a Koch fractal. This fixed
boundary condition simulates a non-deformable substrate. The sample is
compressed from above by a rigid rectangular indenter of length $r$,
centrally positioned on the top surface, with an indentation given by
$\delta$. As depicted in Fig.~\ref{fig1}, at $n=0$, the sample
maintains a rectangular shape with height $H$ and length $L$. However,
the fractal may grow either downwards (in the same direction as the
applied stress) or upwards (in the opposite direction), altering the
sample's mass by adding or removing material, respectively. Down-grown
and up-grown fractals for $n$ up to 4 are displayed, and we examine
both scenarios. Only the bottom surface exhibits roughness, while the
top remains flat to avoid numerical issues at the contact
interface. Our simulations utilize parameters typical of
Polyacrylamide gels, with $E_s =10^3$ Pa, $\rho=10^3$ kg/m$^3$,
$\nu=0.49$, representing materials with elastic moduli ranging from
hundreds to thousands of Pascal, by changing percentages of acrylamide
and bis-acrylamide in PBS (phosphate-buffered saline)
solution~\cite{Caporizzo2015, Tse2010}.

The elastodynamics equations can effectively describe the deformation
of a continuum elastic medium, which provides a comprehensive
framework for understanding momentum conservation and the relationship
between strain and stress~\cite{Spencer2004, Zielinski2013}. These
equations are formulated as
\begin{eqnarray} \label{ref:elastodinamica} \nonumber
\nabla \cdot \mathbf{\sigma} + \mathbf{f} &=& \rho
\ddot{\mathbf{u}},\\ \nonumber \mathbf{\varepsilon} &=& \frac{1}{2}
\left[ (\nabla \mathbf{u})^T + \nabla \mathbf{u} + (\nabla
  \mathbf{u})^T \nabla \mathbf{u} \right],
\end{eqnarray}
where $\mathbf{\sigma}$ represents the Cauchy stress tensor,
$\mathbf{\varepsilon}$ the strain tensor, $\mathbf{u}$ the
displacement vector, and $\mathbf{f}$ the volumetric forces. The
stress and strain tensors are interconnected through the material's
constitutive relation, which, for small deformations, can be
simplified by the generalized Hooke's law $\mathbf{\sigma} =
\mathbf{C}:\mathbf{\varepsilon}$, where the symbol $:$ denotes the
tensorial inner product, and the fourth-order stiffness tensor is
given by
\begin{equation}
\mathbf{C}
= \left[ \begin{array}{ccccccc}
\lambda + 2\mu & \lambda & \lambda  & 0 & 0 & 0 \\ \lambda & \lambda + 2\mu & \lambda & 0 & 0 & 0 \\ \lambda  & \lambda & \lambda + 2\mu & 0 & 0 & 0 \\ 
0 & 0 & 0 & \mu & 0 & 0 \\
0 & 0 & 0 & 0 & \mu & 0 \\
0 & 0 & 0 & 0 & 0 & \mu \\
\end{array} \right],
\end{equation}
where $\lambda=\nu E/[(1+\nu)(1-2\nu)]$ and $\mu=E/[2(1+\nu)]$ are,
respectively, the Lamé's first and second parameters.

\begin{figure}[t]
  \centering
  \includegraphics[width=0.9\textwidth]{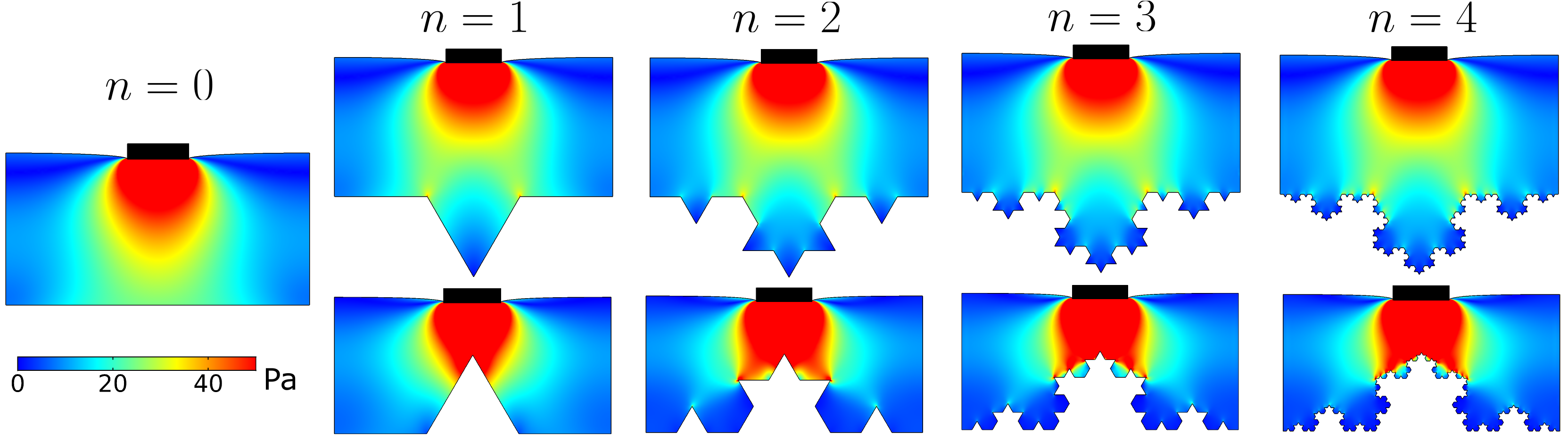}
  \caption{The stress field for up to the fourth generation in both
    \textit{up} and \textit{down} fractal cases for $H=0.5$ cm and
    $L=1.0$ cm. The color gradient represents the Cauchy stress
    magnitude, measured in Pascals (Pa).} \label{fig2}
\end{figure}

In our analyses, we apply indentation up to 4\% of $H$, leading to
stationary conditions where $\ddot{\mathbf{u}} = 0$ and $(\nabla
\mathbf{u})^T \nabla \mathbf{u} =0$, with no external volumetric
forces, $\mathbf{f}=0$. The bottom surface is immovable due to fixed
boundary conditions ($\mathbf{u}=0$).  These conditions simplify the
equations to
\begin{eqnarray}
    \nabla \cdot \mathbf{\sigma} &=& 0,\\
    \mathbf{\varepsilon} &=& \frac{1}{2} \left[ (\nabla \mathbf{u})^T +
      \nabla \mathbf{u} \right],
\end{eqnarray}
commonly known as the compatibility equations. These equations are
solved using the commercial software \textit{COMSOL Multiphysics},
which employs the finite element method to manage the computational
complexities~\cite{Comsol}. The results,
including stress and strain distributions and the contact force,
$F_c$, between the sample and the indenter for a given indentation
$\delta$, are calculated and visually represented. The numerical mesh
used for these calculations is shown in Fig.~\ref{fig1} for $n=4$.

\section{Results}

The geometry of our 2-D model is straightforward, comprising a
rectangular sample in contact with a similarly shaped rectangular
indenter. One of the most fascinating aspects of our model is the
incorporation of a rough surface in the form of a fractal generation
at its base. This distinctive feature is a critical element of our
study. It allows us to explore the effects of self-similarity across
multiple scales, and more importantly, it alters the stress
distribution within the material. Fig.~\ref{fig2} provides a detailed
view of the stress field distribution within the sample across various
fractal generations for both \textit{up} and \textit{down}
geometries. As expected, for a surface without fractal growth ($n =
0$), the stress field disperses uniformly from the indenter's contact
region throughout the sample. However, as $n$ increases, the stress
field becomes concentrated at pointed regions on the fractal surface,
disrupting the uniform distribution. This effect is further
accentuated in the upward fractal orientation, where high-stress
concentrations are observed on the bottom surface, as indicated by the
red regions in the figure. Although measuring stress maps in bulk
samples is challenging, experimental techniques can be applied to
obtain such information on surfaces~\cite{Suki2017}, allowing for a
potential comparison with our results.

The Hertz model for mechanical contact provides critical insights into
the indentation of purely elastic materials. Using a rigid flat-ended
cylindrical indenter results in a linear increase in contact force
with indentation depth $\delta$~\cite{Sneddon1965, Sousa2021}, as
expressed by
\begin{equation}
   F_c(\delta) = \frac{2r}{(1-\nu^2)} E\delta,
   \label{ref: eqHertz}
\end{equation}
where $E$ represents the effective Young's modulus. This relationship
directly reflects Hooke's Law, where the stiffness constant, $k
=\frac{2r}{(1-\nu^2)} E$, can be calculated from the slope of the
contact force curve. We show that $E$ depends on $n$, indicating a
divergence from the actual Young's modulus of the sample, $E_s$.

\begin{figure}[t]
  \centering
  \includegraphics[width=0.8\columnwidth]{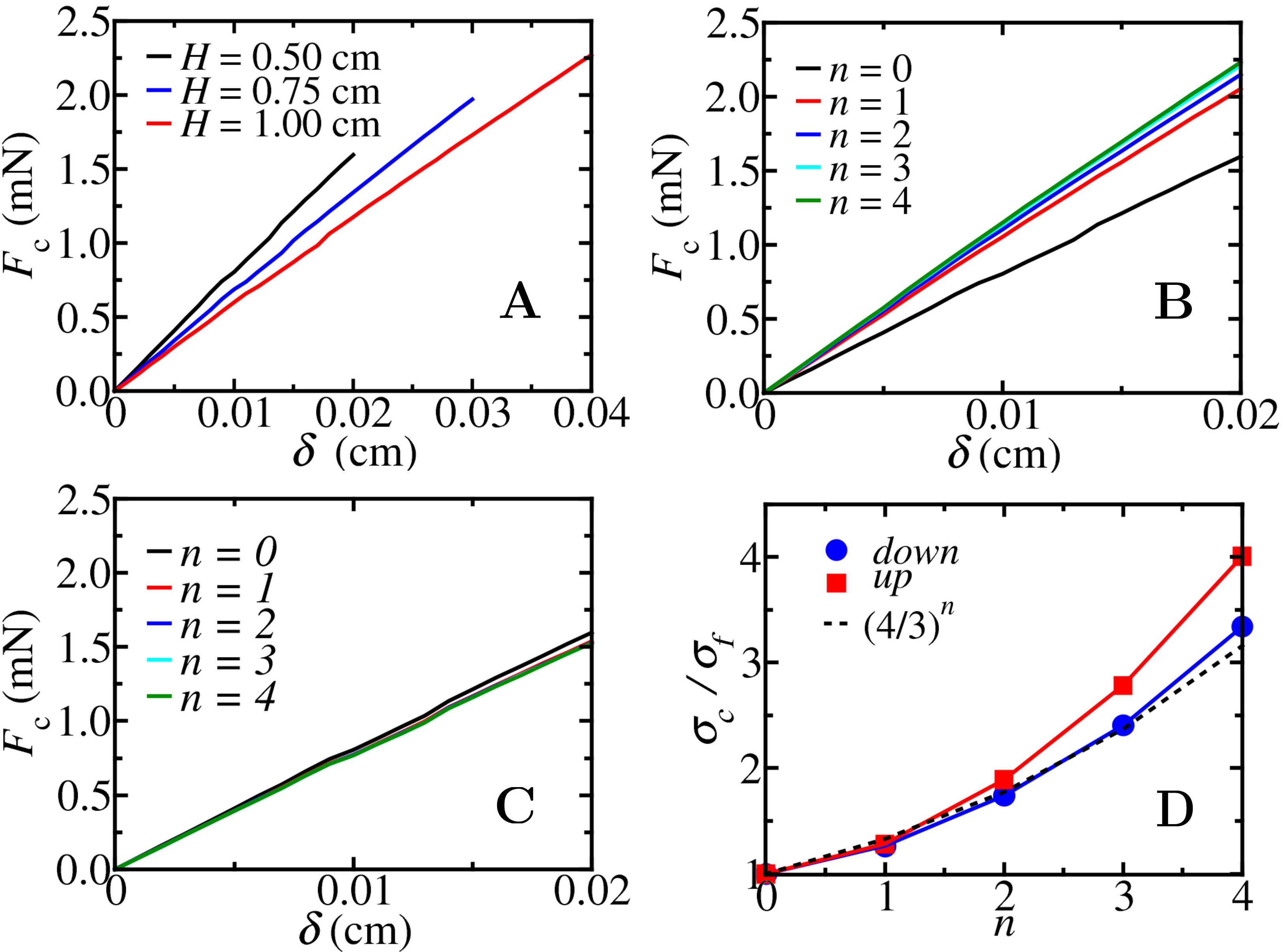}
  \caption{(A) Contact force curve versus indentation for different values of $H$ in the flat sample ($n=0$). The slope of these curves reflects the sample's rigidity. Frames (B) and (C) show similar curves to (A) but for a fixed $H=0.5$ cm and several fractal generations ($n$) in the \textit{up} and \textit{down} fractal scenarios, respectively. These plots effectively demonstrate the influence of fractal geometry on the force-indentation relationship, with significant differences observed in comparison to the flat sample. Additionally, frame (D) depicts the ratio of the stress exerted on the contact surface, $\sigma_c$, to that on the fractal surface, $\sigma_f$, as the fractal generation increases. This graph shows us the accumulation of the stress through the elastic sample. Both \textit{down} and \textit{up} fractal cases slightly diverge from the growth behavior of the Koch fractal, $(4/3)^n$, as $n$ increases.}  \label{fig3}
\end{figure}

Figure \ref{fig3}(A) displays the Hertz curve for the flat case
($n=0$), showing that the spring constant decreases with increasing
sample height, consistent with finite-size
effects~\cite{Costa2022}. The influence of fractal geometry on these
measurements is evident in the force curves in
Figs.~\ref{fig3}(B) and (C). Fractal generation slightly reduces stiffness,
with a more significant impact in the upward scenario. By analyzing
the slope of these curves, we can infer changes in stiffness,
illustrating how fractal geometry influences the material's response
to indentation. These results suggest that as fractal generation $n$
increases, the stiffness modulus $E$ tends towards a specific value
significantly influenced by the sample height.

To calculate the ratio between the average stress on the contact
surface, $\sigma_c$, and that on the fractal surfaces, $\sigma_f$, we
assume the stress is uniformly transmitted along the vertical
direction through the sample. In this scenario, the disparity between
the stresses should compensate for the difference in the perimeter of
each surface. Thus, $\sigma_f \propto 1/P(n)$, where $P(n)=(4/3)^n L$
is the perimeter of the $n^{th}$ generation of the Koch fractal
initial length $L$~\cite{Feder2013}. Similarly, $\sigma_c \propto
1/S(\delta)$, where $S(\delta)$ is the actual length on the top
surface, which depends on the indentation due to its
deformability. For small indentations, $S(\delta) \approx
L$. Therefore, the ratio between the stresses can be computed as
\begin{equation}
    \frac{\sigma_c}{\sigma_f} \propto \frac{P(n)}{S(\delta)} \propto \bigg(\frac{4}{3}\bigg)^{n} 
    \frac{L}{S(\delta)} \approx \bigg(\frac{4}{3}\bigg)^{n},
\end{equation}
yielding the same growth rule as the Koch fractal. However, since the
stress is not uniformly distributed along the sample, the ratio
$\sigma_c/\sigma_f$ as a function of the fractal generation $n$, shown
in Fig.~\ref{fig3}(D), slightly diverges from this prediction. The
divergence observed in the simulations, both in the \textit{up} and
\textit{down} fractal cases, from the $(4/3)^n$ behavior highlights
the non-uniformity of the stress propagation due to the emergence of
stress concentration areas with increasing $n$.

\begin{figure}[t]
  \centering
  \includegraphics[width=0.9\columnwidth]{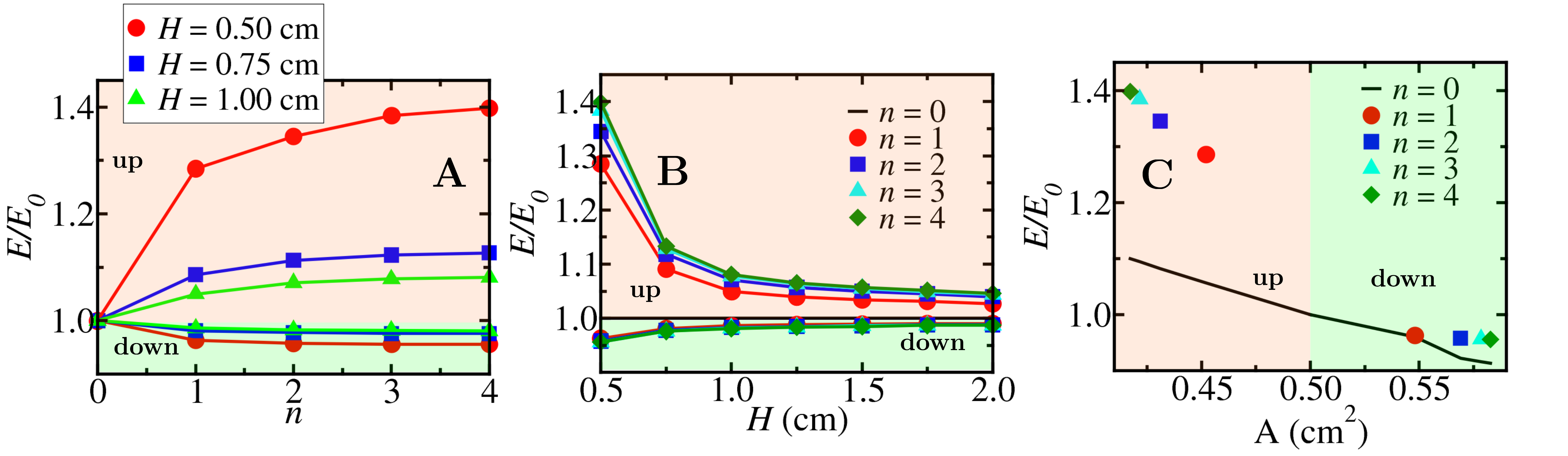}
  \caption{Young's modulus, $E(H,n)$, measured in a sample with height
    $H$ and generation $n$, normalized to the value of $E$ with the
    same $H$ but $n=0$, which we call $E_0$, in both \textit{down} and
    \textit{up} fractal orientations. (A) shows how $E/E_0$ changes
    with $n$ for different values of $H$ while (B) shows changing with
    $H$ for different values of $n$. In frame (C), we adjust $H$ to
    compensate for the area lost or gained due to fractal growth,
    keeping the sample's area constant and equal to $HL$, the area
    without fractal boundary. } \label{fig4}
\end{figure}

\begin{figure}[t]
  \centering
  \includegraphics[width=0.9\columnwidth]{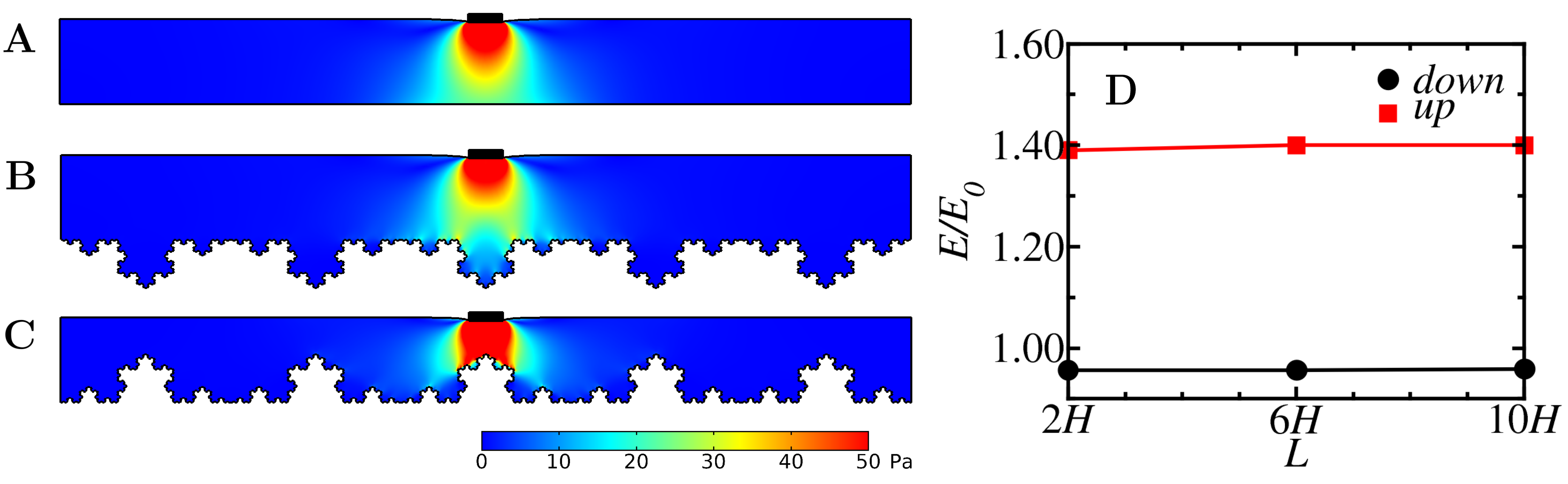}
  \caption{The magnitude of the Cauchy stress, measured in Pascals (Pa), is represented by a color gradient for $H = 0.5$ cm and $L=10H$. The sample without fractal ($n=0$) is shown in (A), while the fractal cases \textit{up} in (B) and \textit{down} (C) for $n=4$. The ratio $E/E_0$, defined in Fig.~\ref{fig4}, is maintained regardless of the length $L$.} \label{fig5}
\end{figure}

The graphs in Fig.~\ref{fig4} present the effective Young's modulus
normalized by the value, $E_0$ where no fractal is grown with height
$H$, for both \textit{down} and \textit{up} cases. By normalizing
Young's modulus, we can track the percentage changes as fractal
generation progresses compared to a smooth sample. Fig.~\ref{fig4}(A)
demonstrates that fractal surfaces can decrease or increase the
measured elastic modulus. In the \textit{down} fractal scenario,
reductions are generally below 5\%, while in the \textit{up} fractal
scenario, increases can reach 40\%. As expected, $E/E_0$ approaches
one as $H$ increases, as shown in frame (B). As $H$ increases, the
influence of surface roughness diminishes, with both \textit{up} and
\textit{down} fractals converging to typical values, illustrating the
lesser impact of fractal features on larger samples. Increasing $H$
leads to a slow convergence of the analytical Young's modulus to the
value input in the numerical model. In Fig.~\ref{fig4}(C), the fractal
was generated at a specific $n$ while $H$ was adjusted to conserve the
same area of $n=0$ (equal to $HL$). The solid black line represents
the expected behavior for flat samples. Even with the area adjustment,
the \textit{up} fractal configuration maintains a relatively high
Young's modulus. These graphs effectively demonstrate the substantial
impact of fractal generation and configuration on the elastic
properties, underscoring how fractal dimensions influence material
behavior under mechanical stress.
We also analyze different cases with varying $L$ ($L=2H$, $6H$, $10H$) for both \textit{up} and \textit{down} scenarios, as depicted in Fig.~\ref{fig5}(A-C) for $L=10H$ and $n=4$. Normalization shows that the effects of fractal generation remain constant despite the increase in $L$, as demonstrated in Fig.~\ref{fig5}(D).

\section{Conclusions}

Our work reveals the significant impact of surface roughness,
particularly fractal geometry, on the accuracy of indentation
experiments used to measure the elastic properties of materials. We
have demonstrated that the fractal nature of the bottom surface is a
critical factor in scenarios involving thin films or textured
substrates, where it significantly alters the mechanical response
during indentation. The structural complexity introduced by fractal
patterns has essential implications for stress distribution and the
force-indentation relationship. Specifically, in the \textit{down}
fractal scenario, where the material is added, reductions in the
elastic modulus ranged between 2\% and 4\% compared to the case
without a fractal boundary, depending on the sample height
$H$. Conversely, in the \textit{up} fractal scenario, where the
material is removed, increases in the modulus reached up to 40\%. We
also analyzed cases where the amount of material was conserved by
adjusting the height $H$ to maintain the same area as a smooth
sample. Even in these scenarios, the fractal boundary significantly
influenced the elastic measurements, demonstrating that fractal
geometry can lead to deviations from expected mechanical behavior,
regardless of material conservation. Our findings not only highlight
the necessity of accounting for surface roughness, particularly
fractal geometry, in the design and analysis of indentation
experiments, but also point to potential applications in
biophysics. Particularly, our work could contribute to a better
understanding of rheological measurements in living cells. The
influence of fractal-like surface roughness could help explain the
variations observed in the mechanical properties of cells, especially
in the context of cellular adhesion and interactions with substrates
of varying stiffness and roughness. This potential for practical
applications should inspire and encourage further exploration in this
field. While our research is focused on deterministic fractals, it
opens up substantial potential for future studies to explore random
fractals, such as self-affine surfaces or other geometric roughness
models. These investigations could lead to the development of more
effective diagnostic tools and material design strategies in
engineering and medical research.

\begin{acknowledgments}
The authors acknowledge the financial support from the Brazilian
agencies CNPq, CAPES, and FUNCAP.
\end{acknowledgments}

Author Declarations: The authors have no conflicts to disclose


\begin{thebibliography}{100}


\bibitem{Allara2005} D. Allara, A perspective on surfaces and
  interfaces, Nature 437 (2005)
  638. https://doi.org/10.1038/nature04234

\bibitem{Oliveira2004a} C. L. N. Oliveira, J. A. K. Freire,
  V. N. Freire, and G. A. Farias, Effects of interfacial profiles on
  quantum levels in In$_x$Ga$_{1-x}$As/GaAs graded spherical quantum
  dots, App. Surf. Sci. 237 (2004)
  266. https://doi.org/10.1016/j.apsusc.2004.06.048

\bibitem{Oliveira2004b} C. L. N. Oliveira, J. A. K. Freire,
  V. N. Freire, and G. A. Farias, Inhomogeneous broadening arising
  from interface fluctuations in strained In$_x$Ga$_{1-x}$As/GaAs and
  (In$_u$Ga$_{1-u}$As)$_v$ (InP)$_{1-v}$/InP quantum wells,
  App. Surf. Sci. 234 (2004) 38.
  https://doi.org/10.1016/j.apsusc.2004.05.055

\bibitem{Prutton1994} M. Prutton, Introduction to Surface Physics,
  Oxford University Press, 1994.

\bibitem{Andrade2004} J. S. Andrade Jr., E. A. A. Henrique,
  M. P. Almeida, M. H. A. S. Costa, Heat transport through rough
  channels, Phys. A 339 (2004) 296.
  https://doi.org/10.1016/j.physa.2004.03.066

\bibitem{Oliveira2010} C. L. N. Oliveira, F. K. Wittel, J. S. Andrade
  Jr, H. J. Herrmann, Invasion percolation with a hardening interface
  under gravity, Int. J. Mod. Phys. C 21 (2010) 903.
  https://doi.org/10.1142/S0129183110015555

\bibitem{Lenzi2014} E. K. Lenzi, A. A. Tateishi, H. V. Ribeiro, M. K. Lenzi, G. Gonçalves and L. R. da Silva,
Fractional diffusion equation, boundary conditions and surface effects,
J. Stat. Mech. (2014) P08019. https://doi.org/10.1088/1742-5468/2014/08/P08019

\bibitem{Chuah2019} H. G. Chuah, W. H. Tan, B. P. Chang, T. S. Khoo,
  C. Y. Khor, and H. G. How, The influence of surface roughness on
  material dislocation of microindentation using bonded interface
  technique, Tribology - Materials, Surfaces \& Interfaces 13 (4)
  (2019) 191. https://doi.org/10.1080/17515831.2019.1643074

\bibitem{Yang2018} H. B. Yang and M. Dai, Influence of surface
  roughness on the stress field around a nanosized hole with surface
  elasticity, Z. Angew. Math. Phys 69 (2018) 127.
  https://doi.org/10.1007/s00033-018-1022-x

\bibitem{Mohammadi2012} P. Mohammadi and P. Sharma, Atomistic
  elucidation of the effect of surface roughness on
  curvature-dependent surface energy, surface stress, and elasticity,
  Appl. Phys. Lett. 100 (2012)
  133110. https://doi.org/10.1063/1.3695069

\bibitem{McMillan2018} A. McMillan, R. Jones, D. Peng and
  G. A. Chechkin, A computational study of the influence of surface
  roughness on material strength, Meccanica 53 (2018)
  2411. https://doi.org/10.1007/s11012-018-0830-6

\bibitem{Xiao2020} Y. Xiao, L. Wu, J. Luo, and L. Zhou, Mechanical
  response of thin hard coatings under indentation considering rough
  surface and residual stress, Diamond and Related Materials 108
  (2020) 107991. https://doi.org/10.1016/j.diamond.2020.107991

\bibitem{Tiwari2021} A. Tiwari and B. N. J. Persson, Cylinder-flat
  contact mechanics with surface roughness, Tribology Letters 69 (1)
  (2021) 4.  https://doi.org/10.1007/s11249-020-01380-z

\bibitem{Peng2023} G. Peng, Y. Liu, F. Xu, H. Jiang, W. Jiang, and
  T. Zhang, On determination of elastic modulus and indentation
  hardness by instrumented spherical indentation: influence of surface
  roughness and correction method, Mater. Res. Express 10 (8) (2023)
  086503.  https://doi.org/10.1088/2053-1591/acebbb

\bibitem{Tiwari2020} A. Tiwari, A. Almqvist, and B. N. J. Persson,
  Plastic Deformation of Rough Metallic Surfaces, Tribol. Lett. 68
  (2020) 129. https://doi.org/10.1007/s11249-020-01368-9

\bibitem{Sousa2017} J. S. de Sousa, J. A. C. Santos, E. B. Barros,
  L. M. R. Alencar, W. T. Cruz, M. V. Ramos, and J. Mendes Filho,
  Analytical model of atomic-force-microscopy force curves in
  viscoelastic materials exhibiting power law relaxation,
  J. Appl. Phys. 121 (2017) 034901. https://doi.org/10.1063/1.4974043

\bibitem{Sousa2020} J. S. de Sousa, R. S. Freire, F. D. Sousa,
  M. Radmacher, A. F. B. Silva, M. V. Ramos,
  A. C. O. Monteiro-Moreira, F. P. Mesquita, M. E. A. Moraes,
  R. C. Montenegro, and C. L. N. Oliveira, Double power-law
  viscoelastic relaxation of living cells encodes motility trends,
  Sci. Rep. 10 (2020) 4749. https://doi.org/10.1038/s41598-020-61631-w

\bibitem{Lima2024} I. V. M Lima, A. V. S. Silva, F. D. Sousa,
  W. P. Ferreira, R. S. Freire, C. L. N. Oliveira, and J. S. de Sousa,
  Measuring the viscoelastic relaxation function of cells with a
  time-dependent interpretation of the Hertz-Sneddon indentation
  model, Heliyon 10 (2024).
  https://doi.org/10.1016/j.heliyon.2024.e30623

\bibitem{Moura2023} A. L. D. Moura, W. V. Santos, F. D. Sousa,
  R. S. Freire, C. L. N. Oliveira, and J. S. Sousa, Viscoelastic
  relaxation of fibroblasts over stiff polyacrylamide gels by atomic
  force microscopy, Nano Ex. 4 (3) (2023)
  035008. https://doi.org/10.1088/2632-959X/acf1b8

\bibitem{Discher2005} D. E. Discher, P. Janmey, and Y.-L. Wang, Tissue
  cells feel and respond to the stiffness of their substrate, Science
  310 (2005) 1139. https://doi.org/10.1126/science.1116995

\bibitem{Parameswaran2014} H. Parameswaran, K. R. Lutchen, and B. Suki,
A computational model of the response of adherent cells to stretch and changes in substrate stiffness,
J. Appl. Physiol. 116 (2014) 825. doi:10.1152/japplphysiol.00962.2013

\bibitem{Rianna2017} C. Rianna and M. Radmacher, Influence of
  microenvironment topography and stiffness on the mechanics and
  motility of normal and cancer renal cells, Nanoscale 9 (31) (2017)
  11222. https://doi.org/10.1039/C7NR02940C

\bibitem{Wen2011} Q. Wen and P. A. Janmey, Polymer physics of the
  cytoskeleton, Curr. Opin. Solid State and Materials, Science 15 (2)
  (2011) 177.  https://doi.org/10.1016/j.cossms.2011.05.002

\bibitem{Trepat2008} X. Trepat, G. Lenormanda and J. J. Fredberg,
  Universality in cell mechanics, Soft Matter 4 (9) (2008) 1750.
  https://doi.org/10.1039/B804866E

\bibitem{Rebelo2013} L. M. Rebelo, J. S. Sousa, J. Mendes Filho, and
  M. Radmacher, Comparison of the viscoelastic properties of cells
  from different kidney cancer phenotypes measured with atomic force
  microscopy, Nanotechnol. 24 (5) (2013) 055102.
  https://doi.org/10.1088/0957-4484/24/5/055102
  
\bibitem{Lekka2016} M. Lekka, Discrimination between normal and
  cancerous cells using AFM, BioNanoSci 6 (2016)
  65. https://doi.org/10.1007/s12668-016-0191-3

\bibitem{Fisseha2012} D. Fisseha and V. K. Katiyar, Analysis of
  Mechanical Behavior of Red Cell Membrane in Sickle Cell Disease,
  Appl. Math. 2 (2012) 40.  https://doi.org/10.5923/j.am.20120202.08

\bibitem{Abidine2021} Y. Abidine, A. Giannetti, J. Revilloud,
  V. M. Laurent, and C. Verdier, Viscoelastic properties in cancer:
  From cells to spheroids, Cells 10 (7) (2021)
  1704. https://doi.org/10.3390/cells10071704

\bibitem{Bogahawaththa2024} M. Bogahawaththa, D. Mohotti,
  P. J. Hazell, H. Wang, K. Wijesooriya, and C. K. Lee, Energy
  absorption and mechanical performance of 3D printed Menger fractal
  structures, Engineering Structures 305 (2024) 117774.
  https://doi.org/10.1016/j.engstruct.2024.117774

\bibitem{Pavon2021} P. Pavón-Domínguez, G. Portillo-García,
  A. Rincón-Casado, and L. Rodríguez-Parada, Influence of the fractal
  geometry on the mechanical resistance of cantilever beams designed
  through topology optimization, Appl. Sci. 11 (22) (2021) 10554.
  https://doi.org/10.3390/app112210554

\bibitem{Wang2018} J. Wang, Y. Zhang, N. He, and C. H. Wang,
  Crashworthiness behavior of Koch fractal structures, Materials \&
  Design 144 (2018) 229.  https://doi.org/10.1016/j.matdes.2018.02.035

\bibitem{Roy2024} A. Roy and K. Vemaganti, Fractal surface-based
  three-dimensional modeling to study the role of morphology and
  physiology in human skin friction, Surf. Topogr.: Metrol. Prop. 12
  (1) (2024) 015006.  https://doi.org/10.1088/2051-672X/ad1fda

\bibitem{Afferrante2023} L. Afferrante, G. Violano, and G. Carbone,
  Exploring the dynamics of viscoelastic adhesion in rough line
  contacts, Sci. Rep. 13 (1) (2023) 15060.
  https://doi.org/10.1038/s41598-023-39932-7

\bibitem{Sokolov2015} I. Sokolov, Editorial - Fractals: a possible new
  path to diagnose and cure cancer?  Future Oncol. 11 (22) (2015)
  3049. https://doi.org/10.2217/fon.15.211

\bibitem{Baish2000} J. W. Baish and R. K. Jain, Fractals and Cancer,
  Cancer Res. 60 (2000) 3683.

\bibitem{Sedivy1997} R. Sedivy and R. M. Mader, Fractals, Chaos, and
  Cancer: Do They Coincide?  Cancer Investigation 15 (6) (1997) 601.
  https://doi.org/10.3109/07357909709047603

\bibitem{Dokukin2011} M. E. Dokukin, N. V. Guz, R. M. Gaikwad,
  C. D. Woodworth, and I. Sokolov, Cell surface as a fractal: normal
  and cancerous cervical cells demonstrate different fractal behavior
  of surface adhesion maps at the nanoscale, Phys. Rev. Lett 107 (2)
  (2011) 028101.  https://doi.org/10.1103/PhysRevLett.107.028101

\bibitem{Caporizzo2015} M. A. Caporizzo, C. M. Roco, M. C. C. Ferrer,
  M. E. Grady, E. Parrish, D. M. Eckmann, and R. J. Composto,
  Strain-rate Dependence of Elastic Modulus Reveals Silver
  Nanoparticle Induced Cytotoxicity, Nanobiomed. 2 (2015) 9.
  https://doi.org/10.5772/61328

\bibitem{Tse2010} J. R. Tse and A. J. Engler, Preparation of Hydrogel
  Substrates with Tunable Mechanical Properties, Current Protocols in
  Cell Biology 47 (2010) 10.16.1.
  https://doi.org/10.1002/0471143030.cb1016s47

\bibitem{Spencer2004} A. J. M. Spencer, Continuum mechanics, Courier
  Corporation, 2004.

\bibitem{Zielinski2013} R. Zielinski et al., Finite element analysis
  of traction force microscopy: influence of cell mechanics, adhesion,
  and morphology, J. Biomech. Eng. 135 (7) (2013) 071009-1.
  https://doi.org/10.1115/1.4024467

\bibitem{Comsol} COMSOL Multiphysics\textregistered www.comsol.com. COMSOL AB, Stockholm, Sweden.

\bibitem{Suki2017} B. Suki, Y. Hu, N. Murata, J. Imsirovic,
  J. R. Mondoñedo, C. L. N. Oliveira, N. Schaible, P. G. Allen,
  R. Krishnan, E. Bartolák-Suki, A microfluidic chamber-based approach
  to map the shear moduli of vascular cells and other soft materials,
  Sci. Rep. 7 (2017) 2305.  https://doi.org/10.1038/s41598-017-02659-3

\bibitem{Sneddon1965} I. Sneddon, The relation between load and
  penetration in the axisymmetric boussinesq problem for a punch of
  arbitrary profile, Int. J. Eng. Sci. 3 (1965) 47.
  https://doi.org/10.1016/0020-7225(65)90019-4

\bibitem{Sousa2021} F. B. de Sousa, P. K. V. Babu, M. Radmacher,
  C. L. N. Oliveira, J. S. de Sousa, Multiple power-law viscoelastic
  relaxation in time and frequency domains with atomic force
  microscopy, J. Phys. D: Appl. Phys. 54 (2021) 335401.
  https://doi.org/10.1088/1361-6463/ac02fa

\bibitem{Costa2022} D. F. S. Costa, J. L. B. de Araújo,
  C. L. N. Oliveira, J. S. de Sousa, Nanoindentation in finite
  thickness viscoelastic materials, Journal of Applied Physics 132
  (2022) 214701.  https://doi.org/10.1063/5.0127403

\bibitem{Feder2013} J. Feder, Fractals, 5th ed., Springer Science \&
  Business Media, 2013.

\end{thebibliography}
\end{document}